\documentclass[12pt]{article}

\usepackage{epsfig,multicol,multirow,cite}
\usepackage{amsmath,subfigure,latexsym,amssymb}

\def\lsi{\raise0.3ex\hbox{$<$\kern-0.75em\raise-1.1ex\hbox{$\sim$}}}
\def\gsi{\raise0.3ex\hbox{$>$\kern-0.75em\raise-1.1ex\hbox{$\sim$}}}
\def\backder{\raise1.4ex\hbox{$\leftarrow$\kern-0.75em\raise-1.4ex\hbox{$\partial$}}}
\def\forder{\raise1.4ex\hbox{$\rightarrow$\kern-0.75em\raise-1.4ex\hbox{$\partial$}}}

\newcommand{\be}{\begin{equation}}
\newcommand{\ee}{\end{equation}}
\newcommand{\nn}{\nonumber}
\newcommand{\bea}{\begin{eqnarray}}
\newcommand{\eea}{\end{eqnarray}}

\newcommand{\R}{{\kern+.25em\sf{R}\kern-.78em\sf{I} \kern+.78em\kern-.25em}}
\newcommand{\RR}{{\kern+.25em\sf{R}\kern-.6em\sf{I} \kern+.6em\kern-.25em}}
\newcommand{\N}{{\kern+.25em\sf{N}\kern-.78em\sf{I} \kern+.78em\kern-.25em}}
\newcommand{\C}{{\kern+.25em\sf{C}\kern-.50em\sf{I} \kern+.50em\kern-.25em}}

\newcommand{\ri}{{\rm i}}

\makeatletter
\@addtoreset{equation}{section}
\makeatother

\begin{document}

\begin{center}

{\Large\bf Features of a 2d Gauge Theory} \vspace*{4mm} 

{\Large\bf with Vanishing Chiral Condensate} \\

\vspace*{7mm}

David Landa-Marb\'{a}n$^{\rm \, a}$, Wolfgang Bietenholz$^{\rm \, a}$
and Ivan Hip$^{\rm \, b}$

\vspace*{7mm}

$^{\rm \, a}$  Instituto de Ciencias Nucleares \\
Universidad Nacional Aut\'{o}noma de M\'{e}xico \\
A.P. 70-543, C.P. 04510 Distrito Federal, Mexico \\

\vspace*{3mm}

$^{\rm \, b}$ Faculty of Geotechnical Engineering, University of Zagreb \\
Hallerova aleja 7, 42000 Vara\v{z}din, Croatia \\

\vspace*{5mm}

\end{center}

{\em The Schwinger model with $N_{f} \geq 2$ flavors is a simple
example for a fermionic model with zero chiral condensate
$\Sigma$ (in the chiral limit). We consider numerical
data for two light flavors, based on simulations with dynamical
chiral lattice fermions. We test properties and predictions
that were put forward in the recent literature for models
with $\Sigma = 0$, which include IR conformal theories.
In particular we probe the decorrelation of
low lying Dirac eigenvalues, and we discuss the mass anomalous 
dimension and its IR extrapolation. Here we encounter subtleties, 
which may urge caution with analogous efforts
in other models, such as multi-flavor QCD.}

\section{Chiral symmetry and the microscopic \\ Dirac spectrum}

Chiral symmetry plays a key r\^{o}le in our understanding
of systems with light fermions. The chiral condensate 
$\Sigma = - \langle \bar \Psi \Psi \rangle$ is the order 
parameter, which indicates whether this symmetry is 
intact ($\Sigma =0$) or broken ($\Sigma > 0$). The latter is
generic at finite fermion mass $m$, but in the chiral limit $m\to 0$
both scenarios occur, depending on the model and its parameters:\\

$\bullet$ $\Sigma (m \to 0) > 0$ is the familiar situation in QCD
at low temperature, where the $SU(N_{f})_{L} \otimes SU(N_{f})_{R}$
chiral flavor symmetry breaks spontaneously down to
$SU(N_{f})_{L+R}$. In our world we encounter 2 (or 3) light quark flavors 
and quasi-spontaneous chiral symmetry breaking. This gives rise
to 2 (or 8) light pseudo-Nambu-Goldstone bosons, which are identified
with light mesons.

In 2 dimensions, spontaneous symmetry breaking can only occur for 
discrete symmetries, as we know from the Mermin-Wagner Theorem \cite{MW}.
Nevertheless the $N_{f}=1$ Schwinger model \cite{schwingmod}
(Quantum Electrodynamics in 2 space-time dimensions) belongs to 
this class as well, although its chiral symmetry is continuous; 
in this case it breaks explicitly, even at $m=0$, 
due to the axial anomaly. The value $\Sigma (m\to 0) \simeq 0.160 \, g$ 
was predicted theoretically \cite{schwingmod}, and confirmed
numerically \cite{DuHo} ($g$ is the gauge coupling).\\

$\bullet$ The opposite scenario, with $\Sigma (m\to 0) \to 0$,
has recently attracted considerable interest, in particular
because it includes the IR conformal theories.
A vanishing chiral condensate is generally expected at high
temperature, in particular for QCD above the chiral crossover, which 
seems to coincide with the deconfinement phase. It also encompasses
the quenched approximation, and $SU(2)$ gauge fields \cite{Tamas}.

At low temperature, multi-flavor QCD --- in particular the extension
of QCD to $N_{f}= 8$ or $12$ light flavors --- is currently a 
subject of intensive research \cite{Nflarge,Nf12proIRconf,
CHSAoki,Nf12contraIRconf,CHPS}. The question whether or not IR 
conformality emerges --- resp.\ above which number $N_{f}$ this
happens --- is today one of the most controversial issues in the 
lattice community. In particular, for $N_{f}=12$ evidence has 
been reported both for \cite{Nf12proIRconf,CHSAoki,CHPS} and against 
\cite{Nf12contraIRconf} this property.
A prominent motivation is the search for nearly conformal
gauge theories, where the coupling moves only little (``walks'')
in some energy regime, as reviewed in Ref.\ \cite{DDebbio}.
That property is of interest in the framework of the ongoing 
attempts to revitalize technicolor approaches.\\

As a further example of the second scenario, we are going to address 
the $N_{f}=2$ Schwinger model. Its Lagrangian in a continuous 
Euclidean plane reads
\bea
&& \hspace*{-7mm}
{\cal L}(\bar \Psi , \Psi, A_{\mu} ) = \frac{1}{2} F_{\mu \nu} F_{\mu \nu}
+ \nn \\
&&  \hspace*{-7mm} ( \bar \Psi^{(1)},\bar \Psi^{(2)} )
 \left( \begin{array}{cc} \gamma_{\mu} ( 
\ri \partial_{\mu} + g A_{\mu} ) + m & 0 \\
0 &  \gamma_{\mu} ( \ri \partial_{\mu} + g A_{\mu} ) 
+ m \end{array} \right)
\left( \begin{array}{c} \Psi^{(1)} \\ \Psi^{(2)} \end{array} 
\right) \ .  \ \qquad
\label{Lschwing}
\eea
$A_{\mu}(x)$ is an Abelian gauge field $(\mu = 1,2)$, 
and $F_{\mu \nu}$ is the corresponding field strength tensor. 
$\gamma_{\mu}$ are Euclidean Dirac matrices; we can represent them 
by two Pauli matrices. The fermions are given by a 2-component 
spinor field $\Psi^{i}(x)$ for each flavor. Here we consider 
two flavors with degenerate mass $m$. It can be incorporated in
the Lagrangian without breaking gauge symmetry, since this
is a ``vector theory'', where both flavors couple to the gauge 
field in the same way (in contrast to ``chiral gauge theories'', 
such as the electroweak sector of the Standard Model).

For the lattice gauge field we use the standard formulation
in terms of compact link variable $U_{x, \mu} \in U(1)$ (where
$x$ is a lattice site), see {\it e.g.}\ Refs.\ \cite{MM}.
The Grassmann functional integral over the fermion fields yields 
the determinant of the Dirac operator, which the Hybrid Monte
Carlo algorithm deals with \cite{MM}. We will comment on the 
lattice Dirac operator in Section 2.

In this case the coupling $g$ is energy independent, and $N_{f} \geq 2$
is sufficient to attain $\Sigma (m\to 0) \to 0$, as we see from 
the relation \cite{Smilga}
\be  \label{delta}
\Sigma (m) \propto m^{1 / \delta} \ , \qquad
\delta = \frac{N_{f} + 1}{N_{f} - 1} \ ,
\ee
which holds in infinite volume, $V = \infty$. In a finite volume
$ V = L \times L$ --- or when taking the chiral limit and the 
infinite volume limit simultaneously --- the critical exponent 
$\delta$ depends on the dimensionless Hetrick-Hosotani-Iso 
parameter \cite{HHI}
\be  \label{HHIpam}
l = \frac{m}{\pi^{1/4}} \ \sqrt{ 2 L^{3} g} \ .
\ee
Eq.\ (\ref{delta}) holds for $l \gg 1$, whereas the opposite
extreme, $l \ll 1 \ll 2Lg / \sqrt{\pi}\, $, leads to $\delta =1$
(which corresponds to the free fermion \cite{LeuSmi}).\\

The chiral condensate is related to the density $\rho (\lambda )$
of Dirac eigenvalues $\lambda$ at zero by the Banks-Casher
relation \cite{BC},
\be
\frac{1}{\pi} \, \Sigma (m=0) \ = \ ^{\lim}_{\lambda \to 0} \ 
^{\lim}_{m \to 0} \ ^{\lim}_{V \to \infty} \ \rho (\lambda )
\ee
(the order of the limits is specified {\it e.g.}\ in 
Ref.\ \cite{Luescher}). In finite volume, the scenario
of a finite $\Sigma$ implies a plateau of the spectral
density $\rho$ {\em near} $\lambda =0$. 
In the $\epsilon$-regime of QCD, {\it i.e.}\ in a small 4d box, 
the prediction for $\rho (\lambda )$ has been refined by 
Random Matrix Theory \cite{RMT}. The corresponding wiggle structure 
on top of the Banks-Casher plateau agrees with lattice data for
staggered fermions \cite{stag} and for overlap fermions 
\cite{ovRMT,stani}; the latter also capture correctly the dependence 
on the topological sector.

A behavior that corresponds to the $\Sigma = 0$ scenario --- and 
therefore to the {\em absence} of a Banks-Casher plateau --- is a 
power-law for the low-lying Dirac eigenvalue density with some 
exponent $\alpha$,
\be \label{powerlaw}
\rho (\lambda ) = c \, V \, | \lambda |^{\alpha} \ ,
\ee
where $c$ is a constant. In fact,
it is natural to expect $\alpha$ to coincide with the
inverse critical exponent $\delta$, {\it i.e.}\
$\Sigma (m) \propto m^{\alpha}$ \cite{Zwicky}.

In the case of high temperature --- {\it i.e.}\ a short extent in 
Euclidean time --- the factor $V$ in eq.\ (\ref{powerlaw})
represents the {\em spatial} volume, since small 
non-zero Dirac eigenvalues only occur in spatial directions.
This is the scenario studied by T.G.\ Kov\'{a}cs in Ref.\ 
\cite{Tamas}. He postulated for this setting the {\em absence 
of correlations between the Dirac eigenvalues,} {\it i.e.}\ 
a Poisson-type statistics. Thus he assumed
the distribution of small eigenvalues in two disjoint intervals 
to be independent (unlike the Random Matrix behavior).
With the additional assumption (\ref{powerlaw}), he derived the 
first eigenvalue density (for $m=0$) as \cite{Tamas}
\be
\rho_{1} (\lambda) =  c V \lambda^{\alpha} \, 
\exp \Big( -\frac{c V}{\alpha + 1} \lambda^{\alpha + 1} \Big) \ .
\ee
Kov\'{a}cs proceeded from $\rho_{1} (\lambda )$ to $\rho_{2} (\lambda )$
by an integral over the product of the probabilities for having 
a first eigenvalue at $\lambda_{1}$, another one at 
$\lambda > \lambda_{1}$, and no eigenvalue in between.
By iterating this step we obtain
\bea
\rho_{n} (\lambda ) &=& \int_{0}^{\lambda} d \lambda' \
\rho_{n-1}(\lambda ') \, P_{\rm no \, EV} (\lambda' , \lambda )
\, \rho (\lambda) \nn \\
&=& \frac{1}{(n-1)!} \frac{(cV)^{n}}{(\alpha +1)^{n-1}}
\lambda^{n(\alpha +1) -1} \exp \Big( - \frac{cV}{\alpha +1} 
\lambda^{\alpha +1} \Big) \ ,  \label{rhon}
\eea
where the probability for no eigenvalue in some interval
$[ \lambda_{a}, \lambda_{b}]$ is given by
\be
P_{\rm no \, EV} (\lambda_{a} , \lambda_{b} ) =
\exp \Big( \frac{cV}{\alpha +1} ( \lambda_{a}^{\alpha +1}
- \lambda_{b}^{\alpha +1}) \Big) \ .
\ee

\section{Simulations of the 2-flavor Schwinger \\ 
model with chiral fermions}

We are going to confront this prediction with data obtained
in simulations of the $N_{f}=2$ Schwinger model, with dynamical 
overlap hypercube fermions \cite{WBIH,stani}. The latter is a variant
of a Ginsparg-Wilson fermion, where the lattice Dirac operator
is constructed by inserting a truncated perfect hypercube lattice 
Dirac operator $D_{\rm HF}$ into the overlap formula \cite{Neu},
\bea
D_{\rm ovHF}(m) &=& \Big( 1 - \frac{m}{2} \Big) D_{\rm ovHF}(0) + m 
\ , \nn \\
D_{\rm ovHF}(0) &=& 1 + (D_{\rm HF} -1 ) / \sqrt{D^{2}_{\rm HF} -1} \ .
\eea
This provides exact (lattice modified) chiral symmetry
\cite{ML98} at $m=0$, along with an excellent level 
of scaling and locality, as well as approximate rotation symmetry 
\cite{WBIH}. All these properties are far superior to the
standard overlap operator. They are based on the similarity
between the (renormalization group improved) kernel and the 
chiral operator, $D_{\rm HF} \approx D_{\rm ovHF}$. 
Regarding the simulation with a Hybrid Monte 
Carlo algorithm, that similarity enables in addition
the use of a simplified force term \cite{BHSV}.

The simulations were carried out at $\beta = 1/g^{2} = 5$,
which leads to plaquette values close to $0.9\,$. Hence we are dealing
with fine lattices, and a continuum extrapolation is not essential.
The volumes have the shape $L \times L$ with $L = 16 \dots 32$, 
and we consider the light fermion masses $m=0.01$ and $0.06$.
Depending on these parameters, finite size effects may be significant.

We analyze eigenvalues $\lambda_{n}$ of the operator $D_{\rm ovHF}(0)$, 
after mapping them\footnote{We can limit the consideration to
eigenvalues with ${\rm Im} \, \lambda_{n} > 0$; the rest just
supplements a degeneracy factor of 2 after the mapping.}
from the unit circle in the complex plane (with center
and radius 1) onto $\R_{+}\,$, by means of the M\"{o}bius transform
\be  \label{mobi}
\lambda_{n} \to \left| \frac{\lambda_{n}}{1 - \lambda_{n}/2}\right| \ .
\ee
As a generic property, the density of small Dirac eigenvalues depends 
on the topological sector, which can be defined by identifying
the fermion index $\nu$ with the topological charge \cite{Has}.

In a previous consideration with fits to the detailed distributions 
of $\lambda_{1}$, $\lambda_{2}$, $\lambda_{3}$ (and $\lambda_{4}$), 
we obtained good agreement with the exponent $\alpha =3/5$,
in particular in the topologically neutral sector ($\nu =0$) \cite{BHSV}.
On the other hand, in infinite volume one expects $\alpha = 1/3\,$,
based on eq.\ (\ref{delta}). This discrepancy becomes plausible if we
consider the Hetrick-Hosotani-Iso parameter $l$ of eq.\ (\ref{HHIpam}).
In Table \ref{HHItab} we display the values of $l$
in our smallest and largest volume.
\begin{table}[h!]
\centering
\begin{tabular}{|c||c|c|}
\hline
 & $L=16$ & $L=32$ \\
\hline
\hline
$m=0.01$ & $0.455$ & $1.286$ \\ 
$m=0.06$ & $2.728$ & $7.715$ \\
\hline
\end{tabular}
\caption{\it The values of the Hetrick-Hosotani-Iso parameter $l$
(defined in eq.\ (\ref{HHIpam})), for the two fermion masses $m$, and 
the extreme lattices sizes $L$, which we consider in this work.}
\label{HHItab}
\end{table}

\vspace*{-5mm}
\section{Testing the decorrelation of the low-lying Dirac eigenvalues}

\begin{figure}[h!]
\center
\includegraphics[angle=0,width=.75\linewidth]{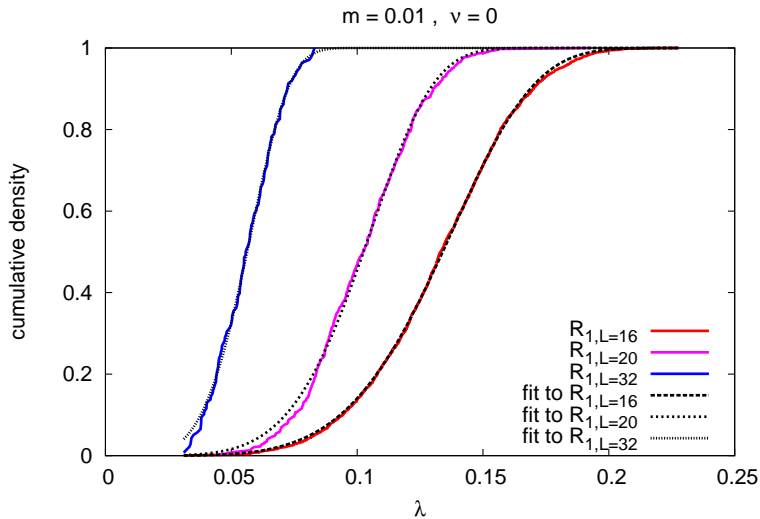}
\caption{\it{The cumulative density of the first Dirac eigenvalue
of the massless operator $D_{\rm ovHF}(0)$,  based on configurations
generated at fermion mass $m=0.01$ and topological charge $\nu =0$, 
on lattices of size $L=16$, $20$ and $32$. In all cases, there is excellent 
agreement between the data and fits to the function $R_{1}( \lambda)$ 
in eq.\ (\ref{Rn}), with adjusted parameters $\alpha$ and $c$.}}
\label{m0.01nu0}
\end{figure}

\begin{figure}[h!]
\center
\includegraphics[angle=0,width=.75\linewidth]{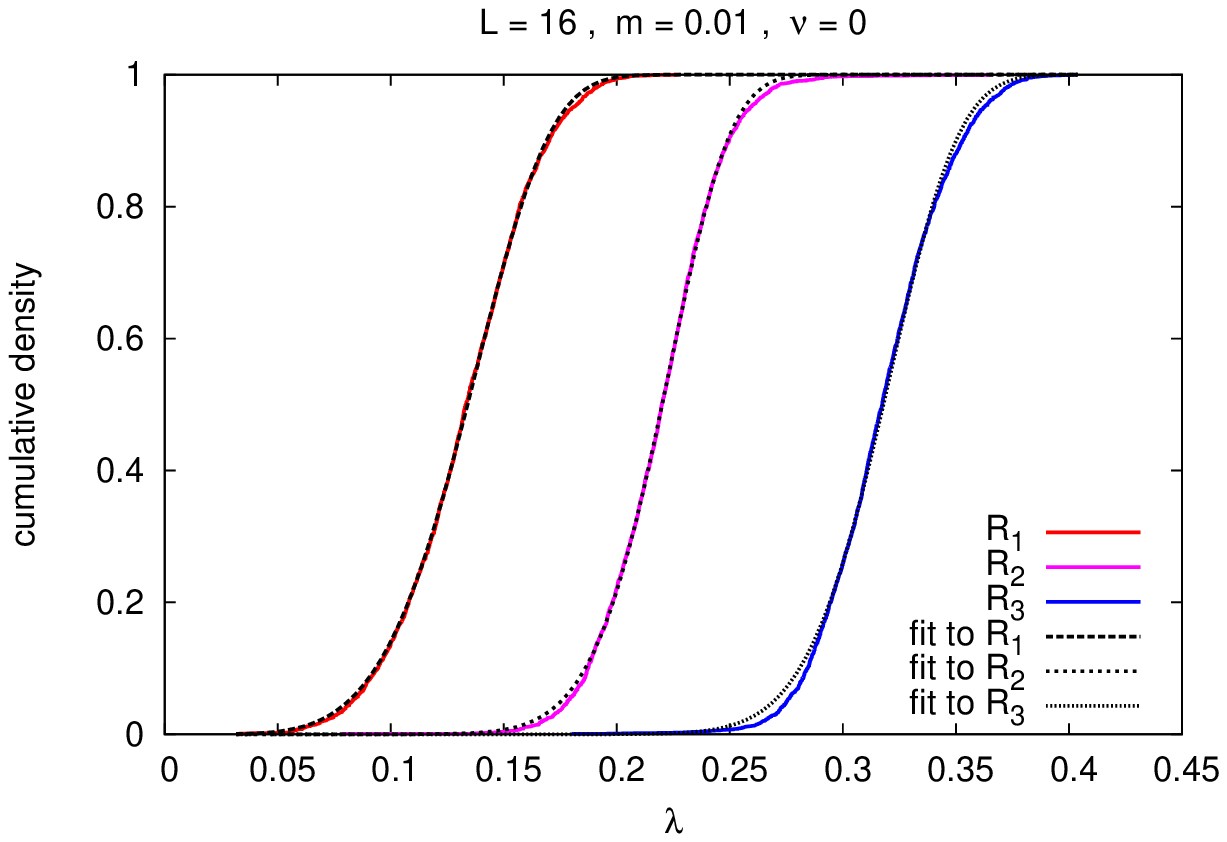}
\includegraphics[angle=0,width=.75\linewidth]{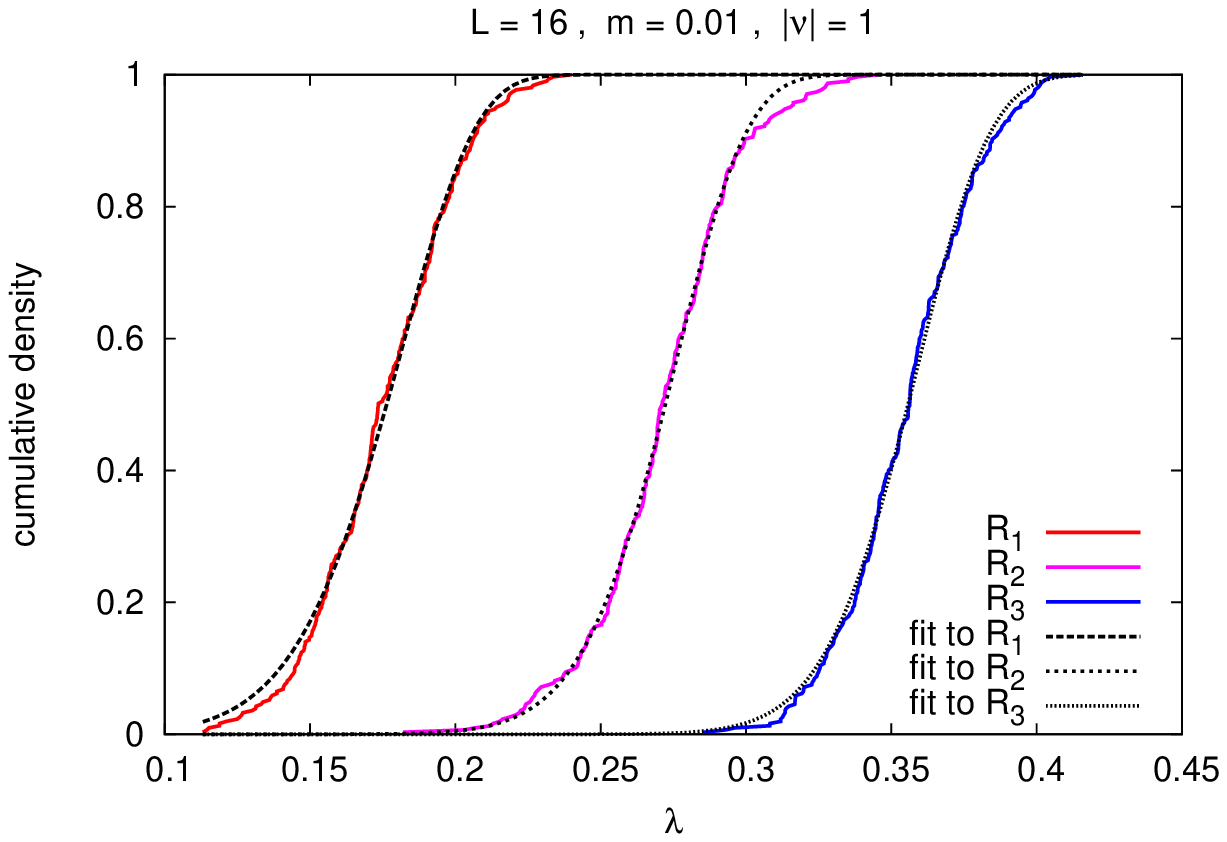}
\caption{\it{Cumulative densities for the first three Dirac eigenvalues,
at $L=16$ and $\nu =0$ (above), $|\nu | =1$ (below). 
In each case, tuning $\alpha$ and $c$ leads to good
agreement with the functions $R_{n}( \lambda)$, $n= 1,2,3$,
in eq.\ (\ref{Rn}). However, the required parameter values
are not consistent, see Table \ref{alphac}.}}
\label{lam123}
\end{figure}

We could test Kov\'{a}cs' conjecture for the model under 
consideration by comparing the functions (\ref{rhon}) to 
histograms. However, in order to avoid the arbitrary choice of a bin 
size, we prefer to compare the corresponding 
{\em cumulative densities,}
\be  \label{Rn}
R_{n}(\lambda ) = \int_{0}^{\lambda} d \lambda ' \, \rho_{n}(\lambda')
= 1 -\exp \Big( - \frac{cV}{\alpha +1} \lambda^{\alpha +1} \Big)
\ \sum_{k=0}^{n-1} \frac{1}{k!} \ \Big( \frac{cV}{\alpha +1}
\lambda ^{\alpha +1} \Big)^{k} \ . \
\ee
Treating the constants $\alpha$ and $c$ as free parameters,
we illustrate in Figure \ref{m0.01nu0} the fits of $R_{n}(\lambda )$
to our data at $m=0.01$ and $L = 16, \ 20$ and $32$,
in the topologically neutral sector ($\nu = 0$).\footnote{The statement 
in the last paragraph of Section 2 is equivalent to our previous 
observation that these distributions collapse onto a single curve 
for all volumes, to quite good accuracy, if the low-lying eigenvalues 
are rescaled as $\lambda_{i} V^{5/8}$. This has been discussed in
Ref.\ \cite{BHSV}, and illustrated there in Figure 11
for $\lambda_{1} \dots \lambda_{4}$, in the 
sectors with $|\nu | = 0 $ and $1$.}
Excellent fits are also achieved if we consider higher eigenvalues,
as Figure \ref{lam123} shows for $R_{n}$, $n = 1,2,3$,
at $m=0.01$, $L=16$, in the sectors $|\nu | =0$ and $1$.\footnote{Note 
that $\lambda_{n}$ refers to the $n$th {\em non-zero} eigenvalue.}

In order to quantify this agreement, Table \ref{KStab1} gives
results of Kolmogorov-Smirnov (KS) test, which compares numerical
data for a cumulative density with a theoretical prediction, see
{\it e.g.}\ Ref.\ \cite{NumRep}. The KS index is between 0 
(extreme disagreement) and 1 (perfect congruousness), and
experience shows that a KS index $\gtrsim 0.5$ characterizes a 
manifestly good agreement. (The low value for
$L=16$, $\nu =0$, $\lambda_{3}$ appears surprising since the
data are not too far from the theoretical curve. However, even
the impact of small deviations is large in this case due to the
high statistics of 2428 configurations.)

\begin{table}[h!]
\vspace*{5mm}
\centering
\begin{tabular}{|c|c|c||c|}
\hline
$L$ & $|\nu |$ & eigenvalue & Kolmogorov-Smirnov index \\
\hline
\hline
16 & 0 & $\lambda_{1}$ & 0.748 \\
\hline
20 & 0 & $\lambda_{1}$ & 0.517 \\
\hline
32 & 0 & $\lambda_{1}$ & 0.962 \\
\hline
\hline
16 & 0 &  $\lambda_{2}$ & 0.648 \\
\hline
16 & 0 &  $\lambda_{3}$ & 0.013 \\
\hline
16 & 1 &  $\lambda_{1}$ & 0.567 \\
\hline
16 & 1 &  $\lambda_{2}$ & 0.727 \\
\hline
16 & 1 &  $\lambda_{3}$ & 0.693 \\
\hline
\end{tabular}
\caption{\it The Kolmogorov-Smirnov confidence index for the
cumulative eigenvalue densities in Figures \ref{m0.01nu0} and
\ref{lam123}, compared with the functions $R_{n}( \lambda)$
in eq.\ (\ref{Rn}), with the optimal parameters $\alpha$ and $c$, 
which are used for the curves in the plots and displayed in 
Table \ref{alphac}.}
\label{KStab1}
\end{table}

\begin{table}[h!]
\centering
\begin{tabular}{|c|c||c|c|c|c|c|c|}
\hline
 $L $ & $| \nu |$ & \multicolumn{2}{|c|}{$n = 1$} & 
\multicolumn{2}{|c|}{$n = 2$} & \multicolumn{2}{|c|}{$n = 3$} \\
\hline
 & & $\alpha$ & $c/1000$ & $\alpha$ & $c/1000$ & $\alpha$ & $c/1000$ \\
\hline
\hline
16 & 0 & 4.199(3) & 0.486(3) & 5.836(5) & $1.40(1)$
& 6.16(1) & 0.274(4) \\
16 & 1 & 7.08(5) & $27(2)$ & 8.08(5) & 
$8.0(6)$ & 8.45(6) & $1.7(1)$ \\
\hline
20 & 0 & 4.23(2) & $1.35(7)$ & 6.00(2) & 9.0(3) & 6.56(3) & 2.6(1) \\
32 & 0 & 3.75(3) & $2.8(3)$  & 5.02(7) & 14(3) & 5.4(1) & 6(2) \\
\hline
\end{tabular}
\caption{\it The parameters $\alpha$ and $c$ (the latter
in units of $10^{3}$), obtained by fitting  formula (\ref{Rn})
to our data at $m=0.01$ for the cumulative densities
of $\lambda_{n}$, $n=1,2,3$.}
\label{alphac}
\end{table}

The corresponding parameters are given in Table \ref{alphac}.
They create first doubt about the confirmation of the
decorrelation property: for fixed $m$, $L$
and $\nu$, the fitting parameters $c$ and $\alpha$ are not 
quite consistent for $R_{1}$, $R_{2}$ and $R_{3}$. 
Of primary interest is the (dimensionless) exponent $\alpha$;
its fluctuation is relatively mild, but all fitted values deviate 
strongly from $\alpha = 3/5$, the value which was determined 
directly from the distributions of these eigenvalues \cite{BHSV}.\\

\begin{figure}[h!]
\center
\includegraphics[angle=0,width=.75\linewidth]{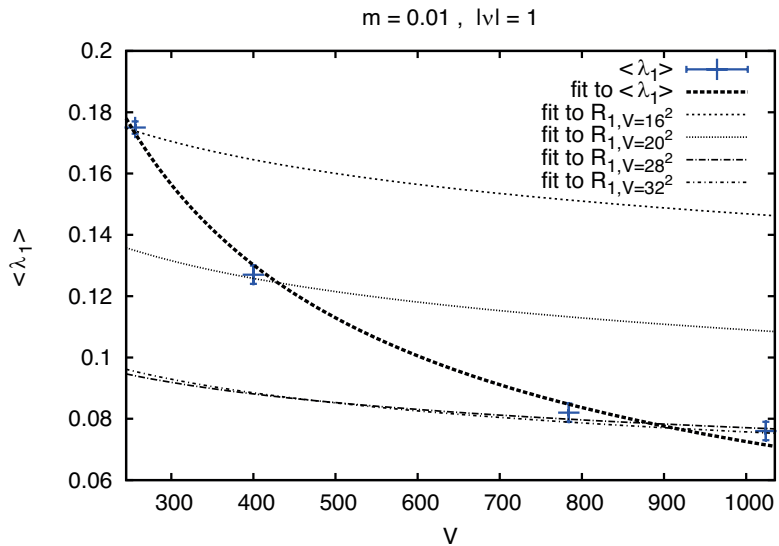}
\caption{\it Fits of the function in eq.\ (\ref{meanlameq})
to our data for $\langle \lambda_{1} \rangle$, at $| \nu | = 1$, 
in volumes $ V = 16^{2},\ 20^{2},\ 28^{2},\ 32^{2}$, with the parameters
of Table \ref{alphac2} (bold line). 
For comparison we show the curves that correspond to the
parameters of Table \ref{alphac}, which are fixed for the
densities of the $\lambda_{1}$ data in separate volumes.}
\label{L16.32}
\end{figure}

Before we continue with the interpretation, we also consider the 
{\em mean eigenvalues.} Formula (\ref{rhon}) predicts them
in terms of $\Gamma$-functions,
\be  \label{meanlameq}
\langle \lambda_{n} \rangle = \int_{0}^{\infty} d \lambda \
\rho_{n}(\lambda ) \, \lambda =
\frac{1}{(n-1)!} \Big( \frac{\alpha +1}{cV} \Big)^{1/(\alpha +1)}
\Gamma \Big( n + \frac{1}{\alpha +1} \Big) \ .
\ee
The corresponding numerical results are given in Table \ref{meanlam}.
\begin{table}[h!]
\centering
\begin{tabular}{|c|c||c|c|c|c|}
\hline
 $L $ & $| \nu |$ & $\langle \lambda_{1} \rangle$ &
 $\langle \lambda_{2} \rangle$ & $\langle \lambda_{3} \rangle$ 
& $\langle \lambda_{4} \rangle$\\
\hline
\hline
16 & 0 & 0.1328(6) & 0.219(1) & 0.3180(6) & 0.3858(5) \\
16 & 1 & 0.175(2)  & 0.271(2) & 0.355(3)  & 0.423(1) \\
20 & 0 & 0.102(2)  & 0.164(2) & 0.238(1)  & 0.294(1) \\
20 & 1 & 0.127(3)  & 0.202(3) & 0.268(2)  & 0.322(1) \\
28 & 1 & 0.082(3)  & 0.132(3) & 0.176(4)  & 0.213(2) \\
32 & 0 & 0.056(3)  & 0.095(4) & 0.133(4)  & 0.165(4) \\
32 & 1 & 0.076(3)  & 0.109(1) & 0.153(3)  & 0.181(3) \\
\hline
\end{tabular}
\caption{\it The mean values of the first four eigenvalues
of the massless Dirac operator, in distinct topological sectors,
for configurations generated at $m=0.01$.}
\label{meanlam}
\end{table}
If we focus on $\langle \lambda_{1} \rangle$, for instance at
$m=0.01$, $|\nu | = 1$ and $V = 16^{2}$, $20^{2}$, $28^{2}$ and $32^{2}$, 
we obtain again a decent fit, see Figure \ref{L16.32} (bold line). 
This is not that conclusive, but not trivial either for four volumes 
and two free parameters.

\begin{figure}[h!]
\center
\includegraphics[angle=0,width=.75\linewidth]{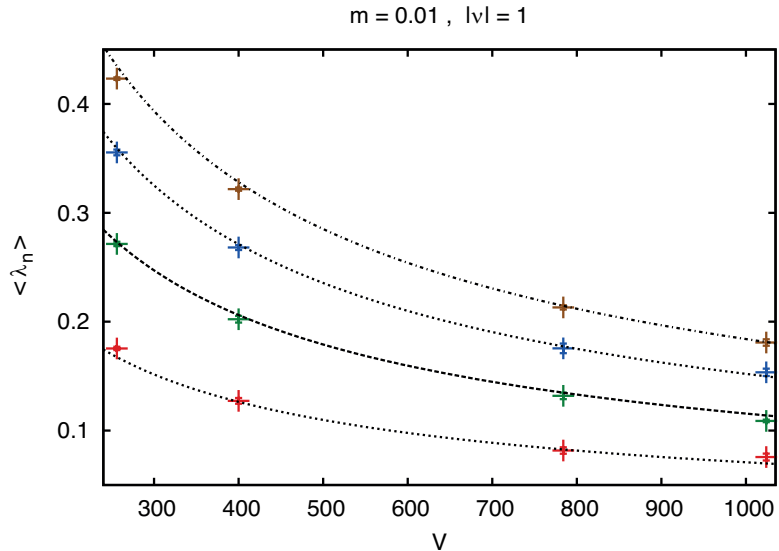}
\caption{\it Mean values of the leading $D_{\rm ovHF}(0)$ eigenvalues
$\langle \lambda_{1}\rangle \dots \langle \lambda_{4} \rangle$
for data obtained at $m=0.01$ and $L=16 \dots 32$. We use one set 
of fitting parameters $\alpha$ and $c$ for $|\nu | =1$.
The four eigenvalues are captured well. However, the parameter values 
--- given in Table \ref{alphac2} --- are incompatible with those 
of the fits in Figures \ref{m0.01nu0} and \ref{lam123}
(given in Table \ref{alphac}).}
\label{lam1to4}
\end{figure}

These fits become highly non-trivial if we extend the consideration
to $\langle \lambda_{n} \rangle$ for $n=1 \dots 4$, and require
a unique set of parameters for each topological sector. The data
in the sector $| \nu | =1$ (where we have results in four volumes)
can be fitted well, see Figure \ref{lam1to4}. The corresponding
parameters are given in Table \ref{alphac2}; they are compatible 
with the value $\alpha = 3/5$, which matches well the detailed
distributions of the leading 3 (or 4) eigenvalues \cite{BHSV}, 
as we mentioned in Section 2.
\begin{table}[h!]
\centering
\begin{tabular}{|c||c|c|}
\hline
$|\nu |$ & $\alpha$ & $c$ \\
\hline
\hline
0 & 0.63(3) & 0.13(1) \\
1 & 0.58(3) & 0.09(3) \\
\hline
\end{tabular}
\caption{\it The parameters $\alpha$ and $c$ obtained by fitting the
mean eigenvalues $\langle \lambda_{1}\rangle \dots \langle \lambda_{4}
\rangle$, at $m=0.01$, in boxes of size $L=16 \dots 32$ 
(cf.\ Figure \ref{lam1to4}).}
\label{alphac2}
\end{table}

If we compare again the required values of $\alpha$ and $c$
for these fits, we see that they differ by orders of magnitudes
from those obtained from the cumulative densities, cf.\ Table \ref{alphac}. 
This is not a contradiction; if we compare the latter values with 
$\langle \lambda_{i} \rangle$ in each single case, it works as
well, as we see from the four finer lines in Figure \ref{L16.32}. 
However, once we fix these values, we cannot capture several volumes.\\

As a final aspect in this context, we consider the {\em unfolded level
spacing density.} One numerates the Dirac eigenvalues of each 
configuration separately in ascending order, puts them all together 
and numerates again. The spacing in this global order between 
eigenvalues, which are adjacent in the ordering of one configuration
--- divided by the number of configurations --- is the unfolded level
spacing $s$. We have shown in Ref.\ \cite{BHSV} that the total spectrum 
follows the statistical distribution of the {\em Chiral Unitary 
Ensemble} \cite{HalVer} (also known as the $\beta=2$ Wigner-Dyson form),
\be
\rho_{\chi {\rm UE}}(s) = \frac{32 s^{2}}{\pi^{2}} \exp (-4 s^{2}/\pi ) \ , 
\ee
as expected.

However, if the microscopic spectrum is decorrelated, the corresponding
unfolded level spacing distribution of eigenvalues near zero 
should approach a {\em Poisson distribution,} 
$\rho_{\rm Poisson} (s) = \exp (- s )$.

In fact, this property has been confirmed for QCD with $2 + 1$ 
light quark flavors above the crossover temperature, by including 
only eigenvalues in the range $0.15 < \lambda < 0.19$ \cite{KovPit}.

\begin{figure}[h!]
\center
\includegraphics[angle=0,width=.5\linewidth]{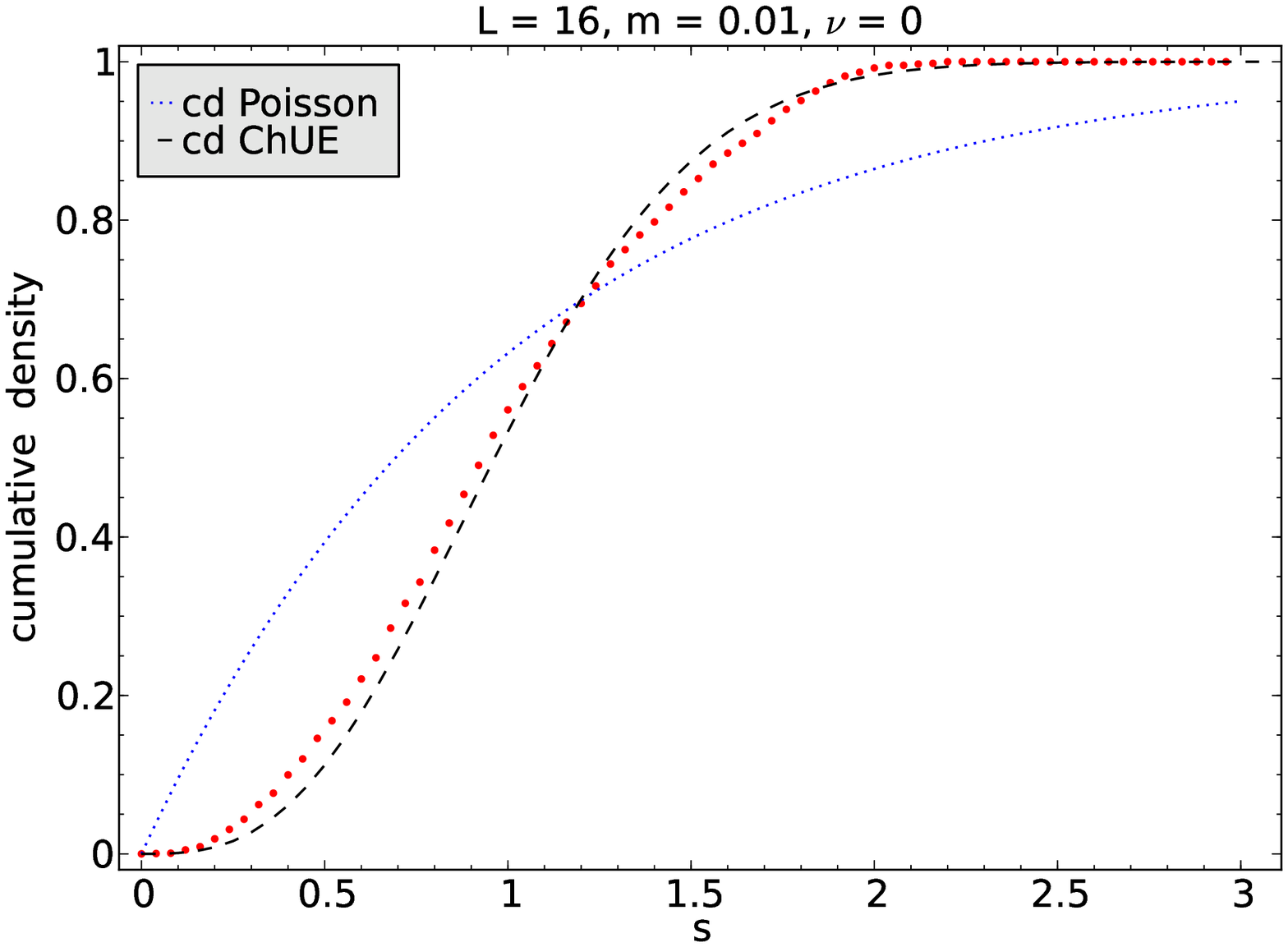}
\includegraphics[angle=0,width=.5\linewidth]{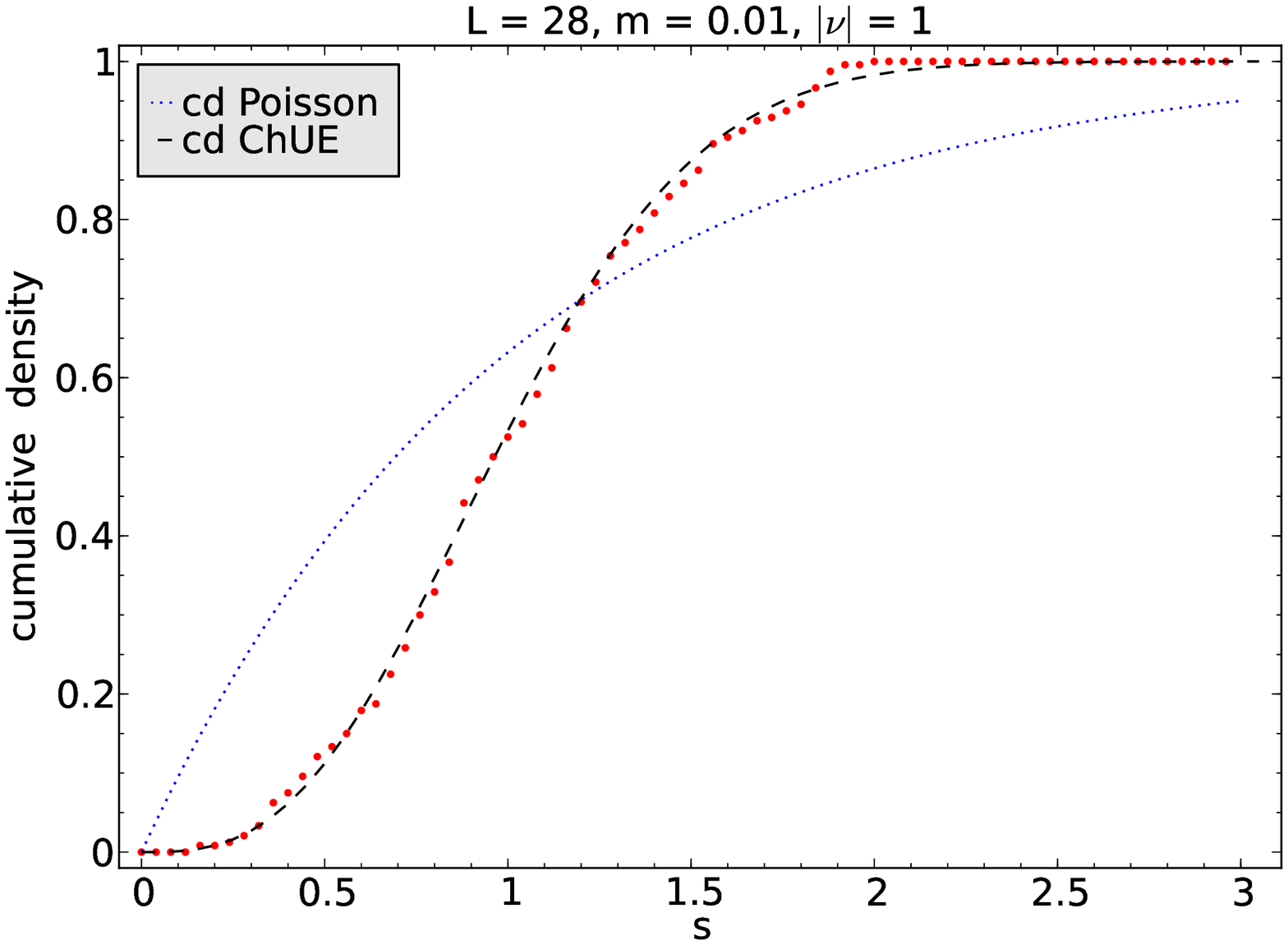}
\includegraphics[angle=0,width=.5\linewidth]{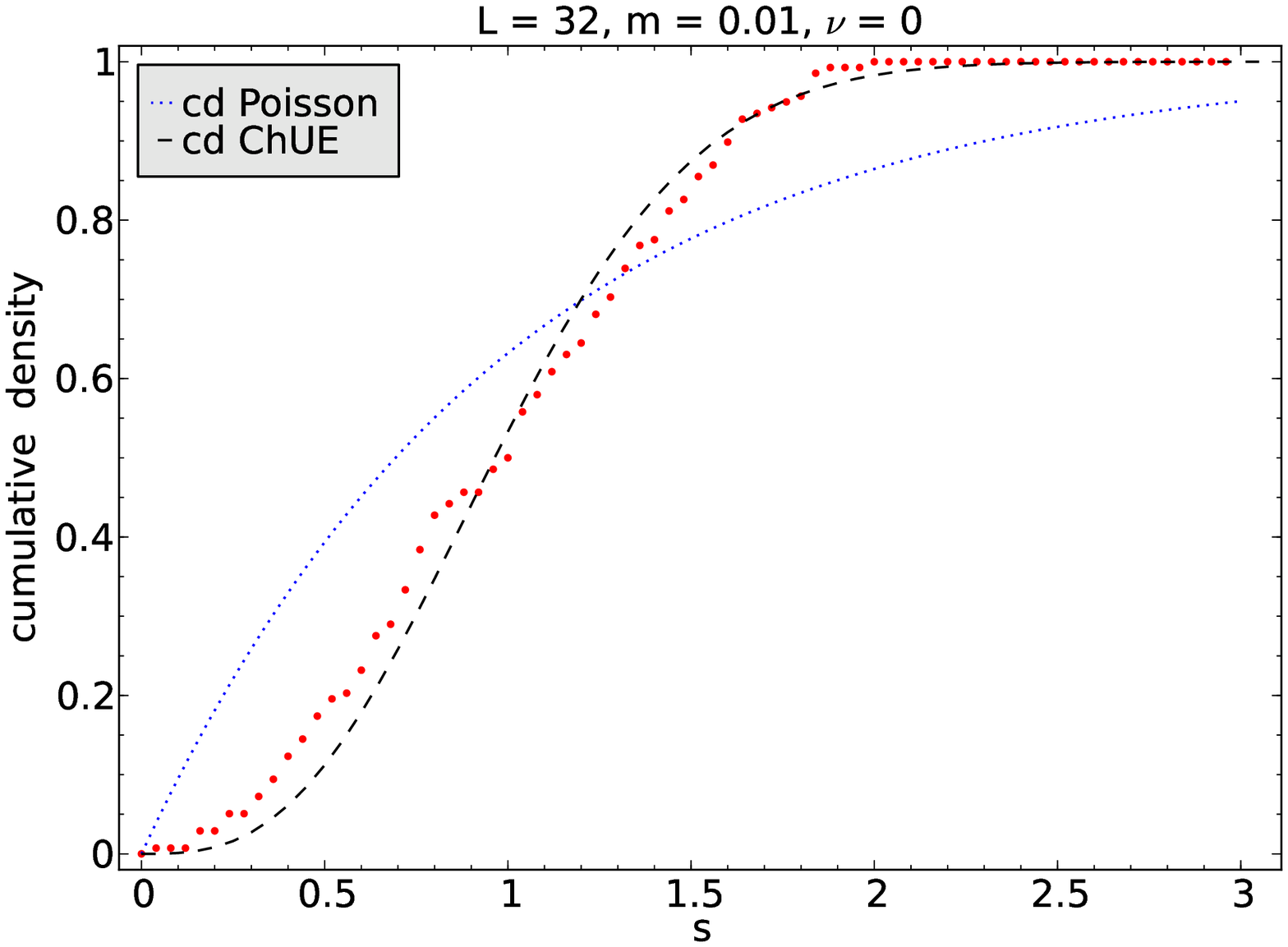}
\caption{\it{The unfolded level spacing density for the
microscopic Dirac spectrum at $m=0.01$ at lattice size 
$L=16, \ 28$ and $32$. We include the lowest two eigenvalues 
of configurations with $\nu = 0$ ($L=16$ and $32$) and 
$|\nu|=1$ ($L=28$). The cumulative densities are in all cases 
close to the Chiral Unitary Ensemble; we do not observe a trend 
towards the Poisson distribution.}}
\label{unfold}
\end{figure}

For our case of the $N_{f}=2$ Schwinger model, three examples for
cumulative densities of the microscopic spectra are shown in 
Figure \ref{unfold}. They are based on the lowest two eigenvalues 
at mass $m=0.01$; in this way we explore the microscopic
regime optimally. In particular we refer to the sector
$\nu = 0$ in sizes $L=16$ and $32$, and to $|\nu |=1$
for $L=28$.

For $L=16$ the statistics is large (2428 configurations),
so we obtain a smooth curve, with a small deviation from 
the Chiral Unitary Ensemble. This is a finite size effect,
which also occurs for the full spectrum at $L=16$, but hardly 
at $L=32$ \cite{BHSV}. The curve for $L=28$ is still quite 
smooth (based on 240 configurations), and in very good
agreement with the Chiral Unitary Ensemble.
The $L=32$ curve is compatible with the same ensemble,
but not that smooth, due to the lower statistics (138 configurations). 
On the other hand, the size $L=32$ and the sector $\nu=0$ 
gives access to smallest eigenvalues, and therefore to the best
probe of the microscopic regime; for the magnitudes
we refer to Table \ref{meanlam}.

In all cases, the densities of $s$ are close to
the distribution of the Chiral Unitary Ensemble, 
even in the microscopic regime that we explore;\footnote{For
$L=16$ we see a small but significant deviation from
$\int_{0}^{s} d s' \, \rho_{\chi {\rm UE}}(s')$, 
which is detected by a tiny KS index
of $5.6 \cdot 10^{-5}$; this is apparently a finite size
effect; for a discussion see Ref.\ \cite{FHLW}.
For $L=28$ the KS index of $0.97$ confirms excellent
agreement, but for $L=32$ it is again reduced to $0.21$, 
though at modest statistics.}
we do not see any trend towards a Poisson distribution.

\section{Mass anomalous dimension}

The numerical measurement of the mass anomalous dimension
is a major issue in the recent lattice literature on possibly
IR conformal theories.

For its evaluation in the $N_{f}=2$ Schwinger model, we follow 
here a procedure which was recently applied in Ref.\ \cite{CHPS}.
Thus we consider the {\em mode number} 
\be
\nu_{\rm mode} (\lambda ) = V \int_{-\lambda}^{\lambda} 
d \lambda ' \ \rho (\lambda ') \ ,
\ee
where $\rho$ is the total Dirac spectral density of $D_{\rm ovHF}(m=0)$.
This quantity --- the cumulative density up to the normalization ---
contains the same information as $\rho (\lambda )$.
It has been studied for $N_{f}=2$ QCD in Ref.\ \cite{Luescher}, 
where also its renormalizability has been demonstrated.

\begin{figure}[h!]
\center
\includegraphics[angle=0,width=.75\linewidth]{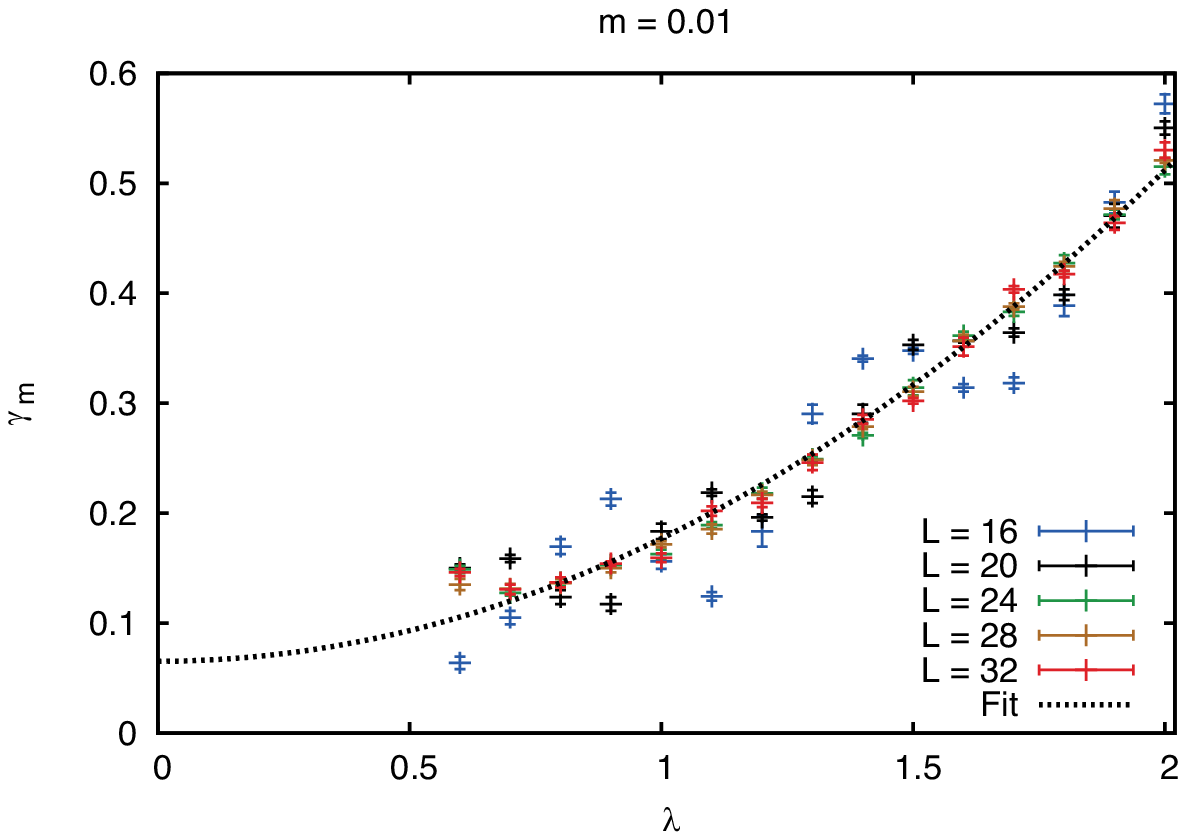}
\vspace*{3mm} \\
\includegraphics[angle=0,width=.75\linewidth]{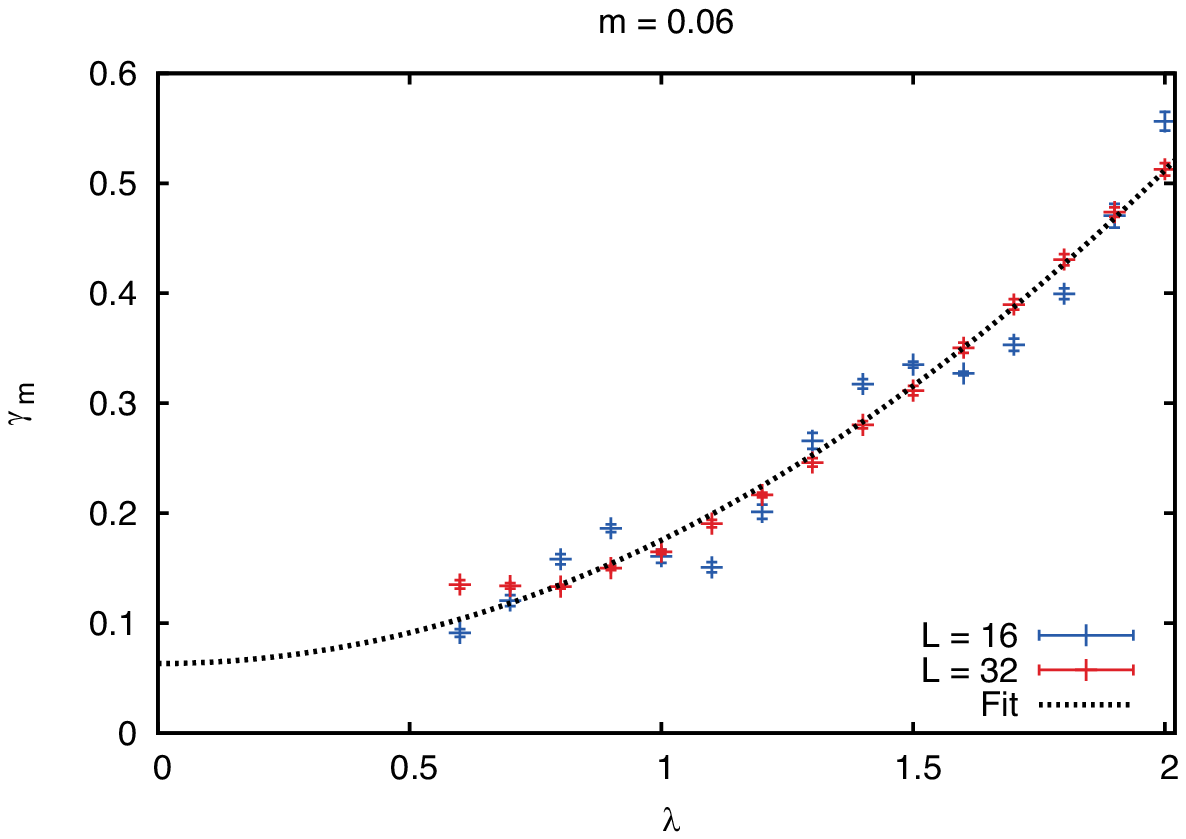}
\caption{\it{The mass anomalous dimension, determined from the
mode number $\nu_{\rm mode}(\lambda )$ in the 
range $\lambda = 0.6 \dots 2$. The results at fermion mass $m=0.01$ 
(above) and $m=0.06$ (below) are very similar; both suggest 
practically the same IR extrapolation to $\gamma_{m}^{*}$,
given in eq.\ (\ref{gammaIR}).}}
\label{massdim}
\end{figure}

If $\rho (\lambda)$ is of the form (\ref{powerlaw}), we obtain
(after mapping the spectrum on $\R_{+}$, cf.\ eq.\ (\ref{mobi}))
\be
\nu_{\rm mode}(\lambda) = \frac{2 c V^{2}}{\alpha +1} \lambda ^{\alpha +1} 
\ .
\ee
By measuring $\nu_{\rm mode}(\lambda)$ we can identify the exponent,
which may be energy dependent, $\alpha (\lambda )$. It is related to the
mass anomalous dimension $\gamma_{m}(\lambda)$ as \cite{Zwicky}
\be
\gamma_{m}(\lambda) = \frac{d}{\alpha (\lambda ) +1 } - 1 \ ,
\ee
where $d$ is the space-time dimension. Free fermions have spectra
$\rho (\lambda ) \propto \lambda^{d-1}$ \cite{LeuSmi}, hence
$\gamma_{m}$ is a measure for the deviation from this behavior
due to interactions. In investigations of candidates for IR conformal 
theories one is most interested in the extrapolation to the IR limit,
which is also our focus,
\be
\gamma_{m}^{*} = \ ^{\lim}_{\lambda \to 0} \ \gamma_{m} (\lambda ) \ .
\ee
Figure \ref{massdim} shows our results for $m=0.01$ and $0.06$ 
and $L=16 \dots 32$. 
For both masses, the data from various volumes agree quite 
well in the range $0.6 \leq \lambda \leq 2$.
This reveals that finite size effects do not affect 
$\gamma_{m} (\lambda \geq 0.6)$ significantly. Moreover, the data
enable a stable IR extrapolation, which agrees very well for both
masses. We infer that, in this framework, the chiral extrapolation is not 
a serious issue either. The two (quadratic) fits in Figure \ref{massdim} 
lead to practically the same IR limit,
\bea
m = 0.01 & ~ : ~ & \gamma_{m}^{*} = 0.065(5) \ , \nn \\
m = 0.06 & ~ : ~ & \gamma_{m}^{*} = 0.063(7) \ .
\label{gammaIR}
\eea

On the other hand, a large Hetrick-Hosotani-Iso parameter, $l \gg 1$,
corresponds to $\alpha =1/3$, as we anticipated in Section 1.
In this limit we obtain $\gamma_{m}^{*} =0.5$. The opposite
limit, $l \ll 1$, leads to $\gamma_{m}^{*} =0$. The value that we
determined from the finite size scaling of the cumulative densities 
$R_{1}$, $R_{2}$, $R_{3}$ in Ref.\ \cite{BHSV}, $\alpha = 3/5$,
corresponds to $\gamma_{m}^{*} =0.25$.
Our fits in Figure \ref{massdim} are based on a regime of higher
energy, so they involve Dirac eigenvalues closer to the bulk.
The corresponding IR extrapolation in eq.\ (\ref{gammaIR}) 
is significantly smaller, and therefore closer to the non-anomalous
value $\gamma_{m}^{*} = 0$ of free fermions.

\section{Conclusions}

We have investigated aspects of the 2-flavor Schwinger
model, as a simple model with $\Sigma = 0$. We first tested
Kov\'{a}cs' conjecture of the decorrelation of low lying
Dirac eigenvalues \cite{Tamas}. The cumulative densities of 
these eigenvalues can be fitted very well to the functions which
correspond to this conjecture. Also the mean eigenvalues
in various volumes can be fitted well to the predicted
form. However, the two fitting parameters take inconsistent
values; in particular the exponent $\alpha$ of eq.\ (\ref{powerlaw})
varies over an order of magnitude for different fits.

As for the unfolded level spacing density, this conjecture
predicts a Poissonian behavior for a restriction to small
Dirac eigenvalues, which turns into the shape of the
Chiral Unitary Ensemble if the full spectrum is included.
However, we did not observe that property either;
as far as we could explore the statistics of the lowest
eigenvalues, their unfolded level spacings are close to
the distribution of the Chiral Unitary Ensemble,
but very far from a Poisson distribution.

Therefore, ultimately the conjecture of low eigenvalue
decorrelation {\em cannot} be confirmed in this model.
On the other hand, this conjecture has been affirmed in the
models studied by Kov\'{a}cs and Pittler \cite{Tamas,KovPit},
which dealt with 4d Yang-Mills gauge theories at high temperature.
This observation is fully consistent with the refined conjecture
that the microscopic eigenvalue decorrelation occurs
if $\Sigma$ vanishes {\em due to high temperature.}
Indeed, according to Ref.\ \cite{localmodes} the inverse
temperature acts as a localization scale for the low lying
Dirac eigenmodes. That scenario includes in particular
QCD above the temperature of the chiral symmetry restoration.

However, this established property left the question open whether
or not the eigenvalue decorrelation also sets in if the chiral 
condensate vanishes for a different reason. Here we
investigated a case where this happens due to a sufficiently
large number of fermion flavors, as it is also expected
in multi-flavor QCD. Contrary to our initial expectation,
the eigenvalue decorrelation conjecture does not lead to a 
consistent picture in this case. Thus our observation restricts
the range of applicability of this interesting conjecture.\\

Regarding the mass anomalous dimension,
this simple model illustrates in a striking manner that the 
determination of $\gamma_{m}^{*}$ is a very subtle issue.
One obtains (apparently) stable results for $\gamma_{m}^{*}$,
which, however, strongly depend on the way how the chiral limit 
and the large volume limit are approached. In general also 
the continuum limit $g \to 0$ is part of the ordering
ambiguity, such that the result for $\gamma_{m}^{*}$
depends on the product $m \sqrt{L^{3} g}$. The formula of 
Ref.\ \cite{Smilga}, eq.\ (\ref{delta}), refers to the procedure 
of taking the continuum and infinite volume limits first, and then
address the chiral condensate at small fermion mass.
However, even if we deal with finite and fixed $g$, $L$ and $m$, 
the outcome for $\gamma_{m}^{*}$ still depends on the energy interval 
that we employ for the IR extrapolation, so this quantity is tricky 
indeed.

This might also provide a hint on why the recent literature
on the corresponding quantity for models with many light quarks 
in $d = 4$, interacting through $SU(3)$ gauge fields, is so 
controversial (cf.\ Section 1), and why it
is particularly hard to determine $\gamma_{m}^{*}$,
see {\it e.g.}\ Refs.\ \cite{CHSAoki,CHPS}. The ongoing discussion 
(and confusion) also includes extensions of QCD 
regarding the number of colors, and quarks in the adjoint or sextet
representation, see Ref.\ \cite{DDebbio} and references therein.
\vspace*{3mm} \\

\noindent
{\bf Acknowledgements:} Stanislav Shcheredin and Jan Volkholz
have contributed to this work at an early stage. We also thank
Poul Damgaard, Stephan D\"{u}rr, Philippe de Forcrand, James Hetrick, 
Christian Hoelbling, Tamas Kov\'{a}cs and Andrei Smilga for helpful 
communication.

This work was supported by the Mexican {\it Consejo Nacional de Ciencia 
y Tecnolog\'{\i}a} (CONACyT) through project 155905/10 ``F\'{\i}sica 
de Part\'{\i}culas por medio de Simulaciones Num\'{e}ricas'', and
by the {\it Croatian Ministry of Science, Education and Sports,} 
project No. 0160013.

\end{document}